\documentclass{ws-procs9x6}

\def\etal {{\it et al.}}
\def\uc#1{\uppercase{#1}}

\begin{document}

\title{ALPHA ANTIHYDROGEN EXPERIMENT}

\author{%
\uc{
M.C.~Fujiwara}$^{a,b*}$,
\uc{
G.B.~Andresen}$^c$,
\uc{
M.D.~Ashkezari}$^d$,
\uc{
M.~Baquero-Ruiz}$^e$,
\uc{
W.~Bertsche}$^f$,
\uc{
C.C.~Bray}$^e$,
\uc{
E.~Butler}$^f$,
\uc{
C.L.~Cesar}$^g$,
\uc{
S.~Chapman}$^e$,
\uc{
M.~Charlton}$^f$,
\uc{
C.L.~Cesar}$^g$,
\uc{
J.~Fajans}$^e$,
\uc{
T.~Friesen}$^b$,
\uc{
D.R.~Gill}$^a$,
\uc{
J.S.~Hangst}$^c$,
\uc{
W.N.~Hardy}$^h$,
\uc{
R.S.~Hayano}$^i$,
\uc{
M.E.~Hayden}$^d$,
\uc{
A.J.~Humphries}$^f$,
\uc{
R.~Hydomako}$^b$,
\uc{
S.~Jonsell}$^j$,
\uc{
L.~Kurchaninov}$^a$,
\uc{
R.~Lambo}$^g$,
\uc{
N.~Madsen}$^f$,
\uc{
S.~Menary}$^k$,
\uc{
P.~Nolan}$^l$,
\uc{
K.~Olchanski}$^a$,
\uc{
A.~Olin}$^a$,
\uc{
A.~Povilus}$^e$,
\uc{
P.~Pusa}$^l$,
\uc{
F.~Robicheaux}$^m$,
\uc{
E.~Sarid}$^n$,
\uc{
D.M.~Silveira}$^o$,
\uc{
C.~So}$^e$,
\uc{
J.W.~Storey}$^a$,
\uc{
R.I.~Thompson}$^b$,
\uc{
D.P.~van der Werf}$^f$,
\uc{
D.~Wilding}$^f$,
\uc{
J.S.~Wurtele}$^e$, AND
\uc{
Y.~Yamazaki}$^o$\\
\uc{
(ALPHA Collaboration)}\\
}

\address{%
$^a$TRIUMF, 4004 Wesbrook Mall, Vancouver BC, V6T 2A3, Canada\\
$^b$Department of Physics \& Astronomy, University of Calgary, AB, T2N 1N4, Canada\\
$^c$Department of Physics \& Astronomy, Aarhus University, DK-8000, Denmark\\
$^d$Department of Physics, Simon Fraser University, Burnaby BC, V5A 1S6, Canada\\
$^e$Department of Physics, University of California, Berkeley, CA 94720-7300, USA\\
$^f$Department of Physics, Swansea University, Swansea SA2 8PP, United Kingdom\\
$^g$Instituto de F\'{i}sica, Universidade Federal do Rio de Janeiro, 21941-972, Brazil\\
$^h$Depart.\ of Physics \& Astronomy, Univ.\ of British Columbia, BC, V6T 1Z4, Canada\\
$^i$Department of Physics, University of Tokyo, Tokyo 113-0033, Japan\\
$^j$Fysikum, Stockholm University, SE-10609, Stockholm, Sweden\\
$^k$Department of Physics \& Astronomy, York University, ON, M3J 1P3, Canada\\
$^l$Department of Physics, University of Liverpool, Liverpool L69 7ZE, United Kingdom\\
$^m$Department of Physics, Auburn University, Auburn, AL 36849-5311, USA\\
$^n$NRCN-Nuclear Research Center Negev, Beer Sheva, IL-84190, Israel\\
$^o$Atomic Physics Laboratory, RIKEN, Saitama 351-0198, Japan\\
$^*$E-mail: Makoto.Fujiwara@triumf.ca
}

\begin{abstract}
ALPHA is an experiment at CERN, whose ultimate goal is to perform a precise test of CPT symmetry with trapped antihydrogen atoms. After reviewing the motivations, we discuss our recent progress toward the initial goal of stable trapping of antihydrogen, with some emphasis on particle detection techniques.
\end{abstract}

\bodymatter

\section{Introduction}

Hydrogen is the most abundant element in the Universe. Experimental and theoretical studies of atomic hydrogen over the past century have helped build the foundation of modern physics. Today, the energy level difference between its 1s and 2s states is measured to a relative precision of 10$^{-14}$, and the ground state hyperfine splitting to 10$^{-12}$, making hydrogen one of the best studied physical systems. On the other hand, the antimatter counterpart of atomic hydrogen, namely antihydrogen, has been only recently produced at low energies~\cite{Nature}. The goal of ALPHA (Antihydrogen Laser Physics Apparatus) is to perform precision tests of CPT via spectroscopic comparisons of hydrogen and antihydrogen atoms. In the longer term, we envision extending our antimatter studies to the gravity sector.

According to the CPT theorem,\cite{CPT} the energy levels of atoms and antiatoms must be identical. Any difference would imply violation of fundamental assumptions in the theorem, which has been proven for pointlike particles in a flat spacetime within the framework of local and relativistic quantum field theory. Precision measurements of antihydrogen atoms thus will potentially confront some of the most fundamental concepts in physics.

In the past decade, Kosteleck\'y and his coworkers have led intensive theoretical investigations on CPT and Lorentz violation.\cite{Datatables} Their model, the so-called Standard-Model Extension (SME), is the most phenomenologically studied theory of CPT and Lorentz violation, and the parameters of the theory have been extensively tested experimentally with laboratory systems using matter particles, as well as astrophysical sources. Yet, no direct comparison of atomic and antiatomic systems\cite{HbarTheory} has been performed to date. Such a measurement will provide a test of CPT and Lorentz violation that is complementary to those using matter-only particles.

One prediction of the SME for hydrogen-antihydrogen comparisons\cite{HbarTheory}  is that for the same relative precision, microwave spectroscopy of hyperfine splitting would give a more sensitive test of CPT violation than laser spectroscopy of the 1s-2s transition. Therefore both types of spectroscopic measurements are worthwhile to be pursued. See Ref.\ \refcite{AIP2008} for a more detailed discussion on fundamental physics motivations for antihydrogen studies.

\section{ALPHA experiment}

Antiatoms as previously produced at CERN, while nearly at rest, were not confined and rapidly annihilated on the walls of the apparatus. In order to probe matter-antimatter symmetry at the highest possible precision, it is essential that the antiatoms be confined in vacuum to allow for detailed interrogation via laser light or microwaves.

In a typical experimental cycle, a beam of $3 \times 10^7$ antiprotons is delivered from the Antiproton Decelerator (AD) every 100 s. Using a pulsed electric field, roughly 50,000 antiprotons with energy less than 3 keV are trapped in the catching trap, where they subsequently cool via Coulomb collisions with a preloaded cold electron plasma. The antiproton-electron mixture is then sympathetically compressed via application of a rotating RF field, and then transferred to the mixing trap. After removal of the electrons, we are left with antiprotons at $\sim$300 K. In parallel, positrons are accumulated and compressed in a buffer gas moderated Penning trap, then transferred to the mixing trap, where they are further cooled and compressed. In this way, $3\times10^4$ antiprotons and $4\times10^6$ positrons are prepared prior to mixing. The two species are then gently mixed by making use of a nonlinear dynamics phenomenon, autoresonance.\cite{AR} If an antihydrogen atom formed during the mixing procedure is cold enough, it will be confined in our Ioffe-type multipolar magnetic trap.\cite{NIM} This magnetic trap confines neutral antiatoms via interaction of the antihydrogen magnetic moment with the magnetic field gradients. The depth of the potential well, which uses state-of-the-art superconducting technology, is limited to $\sim$0.5 K (or 50 $\mu$eV). Given the currently achieved set of parameters, the expected rate for antihydrogen trapping is low. A 3-layer silicon vertex detector, which surrounds the trap region (with a total active area of 8000 cm$^2$), plays a crucial role in indentifying trapped antihydrogen and in rejecting the background. Another novel feature is our ability to shut down the magnetic trap in $\sim$10 ms via a controlled quench of the superconducting magnets. This further reduces the cosmic background via temporal gating.

\section{Recent progress}

Progress towards antihydrogen trapping is faced with many unique challenges associated with the handling of antimatter particles, requiring development of special techniques for particle manipulations. We have made rapid progress since the startup of ALPHA. Our published achievements include: (1) demonstration of trapped plasma stability in a combined  Penning trap (for charged particles) and magnetic trap (for neutral atoms);\cite{Oct} (2) production of antihydrogen in a reduced magnetic field;\cite{1T} (3) development of a technique for antiproton plasma diagnosis based on annihilation detection;\cite{Ramp} (4) sympathetic radial compression of antiproton clouds;\cite{Compression} (5) observation of a new radial transport mechanism, induced by magnetic multipolar fields in a Penning trap;\cite{ZeroRotation} (6) development of antiproton, positron, and electron imaging with a microchannel plate/phosphor detector;\cite{MCP} (7) production (but not yet trapping) of antihydrogen in a multipolar antiatom trap environment.\cite{Formation} A key to the success in achieving these milestones has been the development of sophisticated plasma diagnostic techniques, including antiproton annihilation imaging via the Si vertex detector pioneered in the ATHENA experiment\cite{Imaging,OnOff,Detector}, with which we have unique sensitivity to particle loss processes.

Most recently, in 2009-2010, we made further important steps. First, we demonstrated evaporative cooling of antiprotons clouds to temperatures of order 10 K.\cite{EVC} Evaporative cooling has been used widely for neutral cold atoms in the context of Bose-Einstein condensate studies, but this is the first time it has been accomplished for cold charged ions (with the exception of electron beam ion traps at much higher ($\sim$100 eV) temperatures), let alone for antimatter particles.\cite{Focus}  Second, we have achieved, for the first time, trap conditions and detection sensitivity where observation of antihydrogen trapping could be realistically expected.

We have conducted an extensive search for trapped antihydrogen.\cite{Search} Signatures of annihilations of antihydrogen released from the trap were sought via detection of antiproton annihilations in the Si detector. In order to unambiguously identify rare events against cosmic-ray background, we developed an analysis technique that minimizes experimenter bias. First, event selection criteria (`cuts') were investigated using independent calibration samples without directly analyzing the actual experimental data. (Using the data themselves to optimize the cuts has resulted in numerous instances of experimental bias in the history of particle physics).

Furthermore, our cuts were optimized for the best sensitivity via Monte Carlo pseudo-experiments. Because of the statistical nature of our low event rate experiment, the statistical significance one obtains in a single experiment fluctuates from one experiment to another, according to the Poisson distribution. However, by running a large number of pseudo-experiments, we studied the effects of varying cuts where the results are averaged over a number of trials. Thus, we have derived a set of cuts which would produce a best statistical significance on average.\cite{Richard}.

After these detailed studies of the event selection criteria, the chosen cuts were finally applied to the experimental data. We found 6 events that are consistent with annihilations of trapped antihydrogen atoms. From our cut studies, we estimated our cosmic background to be 0.14 events. Hence, our observation has a significance of 5.6 $\sigma$ against cosmic-ray background. However, there is one other source of potential background, namely antiprotons which could be trapped in our magnetic trap via the magnetic mirror effect. While detailed simulation studies indicated that this possibility was highly unlikely, we could not {\it experimentally} rule out this background. Nonetheless, the trapping conditions and detection sensitivity achieved in these experiments are unprecedented, and observation of candidate events for trapped antihydrogen gives great promise for the techniques developed by ALPHA. The details of the detector analysis will appear in Ref.\ \refcite{Richard}.

\section{Summary and prospects}

Significant progress has been made towards establishing antihydrogen trapping. In the meantime, we are actively preparing for the first spectroscopy on antiatoms via microwaves.\cite{HypInt} Given our high efficiency for antihydrogen annihilation detection, there is a realistic chance that initial spectroscopy measurements could be performed even with a few trapped antiatoms. We are entering a very exciting time for antihydrogen physics.

\section*{Acknowledgments}
This work was supported by CNPq, FINEP/RENAFAE (Brazil), ISF (Israel), MEXT (Japan), FNU (Denmark), VR (Sweden), NSERC, NRC/TRIUMF, AITF (Canada), DOE, NSF (USA), EPSRC and the Leverhulme Trust (UK).


\begin{thebibliography}{xx}

\bibitem{Nature}
M.\ Amoretti \etal,
Nature (London), {\bf 419}, 456 (2002);
G.\ Gabrielse \etal, Phys.\ Rev.\ Lett.\ {\bf 89}, 213401 (2002).

\bibitem{CPT}
G.\ L\"{u}ders, Ann.\ Phys.\ {\bf 2}, 1 (1957).

\bibitem{Datatables}
V.A.\ Kosteleck\'y and N.\ Russell,
arXiv:0801.0287v3.

\bibitem{HbarTheory}
R.\ Bluhm, V.A.\ Kosteleck\'y, and N.\ Russell
Phys.\ Rev.\ Lett.\ {\bf 82}, 2254 (1999).

\bibitem{AIP2008}
M.C.\ Fujiwara \etal, AIP Conf.\ Proc.\ {\bf 1037}, 208 (2008).

\bibitem{AR}
J.\ Fajans and L.\ Friedland,  Am.\ J.\ Phys.\ {\bf 69}, 1096 (2001).

\bibitem{NIM}
W.\ Bertsche \etal, Nucl.\ Instrum. Meth.\ A {\bf 566}, 746 (2006).

\bibitem{Oct}
G.\ Andresen \etal, Phys.\ Rev.\ Lett.\ {\bf 98}, 023402 (2007).

\bibitem{1T}
G.\ Andresen \etal, J.\ Phys.\ B {\bf 41}, 011001 (2008).

\bibitem{Ramp}
G.\ Andresen \etal,
Phys.\ Plasmas {\bf 15}, 032107 (2008).

\bibitem{Compression}
G.\ Andresen \etal,
Phys.\ Rev.\ Lett.\ {\bf 100}, 203401 (2008).

\bibitem{ZeroRotation}
G.\ Andresen \etal,
Phys.\ Plasmas {\bf 16}, 100702 (2009).

\bibitem{MCP}
G.\ Andresen \etal, Rev.\ Sci.\ Inst.\ {\bf 80}, 123701 (2009).

\bibitem{Formation}
G.\ Andresen \etal,
Phys.\ Lett.\ B {\bf 685}, 141 (2010).

\bibitem{Imaging}
M.C.\ Fujiwara \etal, Phys.\ Rev.\ Lett.\ {\bf 92}, 065005 (2004).

\bibitem{OnOff}
M.C.\ Fujiwara \etal, Phys.\ Rev.\ Lett.\ {\bf 101}, 053401 (2008).

\bibitem{Detector}
M.C.\ Fujiwara, AIP Conf.\ Proc.\ {\bf 793}, 111 (2005).

\bibitem{EVC}
G.\ Andresen \etal,
Phys.\ Rev.\ Lett.\ {\bf 105}, 013003 (2010).

\bibitem{Focus}
For a nontechnical description, see http://focus.aps.org/story/v26/st1.

\bibitem{Search}
G.\ Andresen \etal, submitted for publication (2010).

\bibitem{Richard}
R.\ Hydomako, Ph.D.\ thesis, University of Calgary, in preparation.

\bibitem{HypInt}
M.C.\ Fujiwara \etal, Hyperfine Interact.\ {\bf 172}, 81 (2006).

\end{thebibliography}
\end{document}